\documentclass[twocolumn,amsmath,amssymb,prb,floatfix]{revtex4-1}

\usepackage{amsmath}
\usepackage{amsfonts}
\usepackage{bm}
\usepackage{dsfont}
\usepackage{graphicx}
\usepackage{subcaption}
\usepackage{braket}
\usepackage{multirow}

\hyphenchar\font=-1

\usepackage{graphicx}
\usepackage{dcolumn}
\usepackage{bm}
\usepackage{color}
\usepackage{braket}
\usepackage{booktabs}
\usepackage{multirow}
\usepackage{subcaption}

\newcommand{\Id}{\mathbf{I}}
\newcommand{\Hr}{\mathbf{H}}

\newcommand{\Sr}{\mathbf{S}}

\begin{document}

\title{A deep neural network for molecular wave functions in quasi-atomic minimal basis representation}

\author{M. Gastegger}
\email{michael.gastegger@tu-berlin.de}
\affiliation{Machine Learning Group, Technische Universität Berlin, 10587 Berlin, Germany}
\author{A. McSloy}
\affiliation{Department of Chemistry, University of Warwick, Gibbet Hill Road, CV4 7AL Coventry, UK}
\author{M. Luya}%
\affiliation{Department of Chemistry, University of Warwick, Gibbet Hill Road, CV4 7AL Coventry, UK}
\author{K. T. Sch\"utt}
\affiliation{Machine Learning Group, Technische Universität Berlin, 10587 Berlin, Germany}
\author{R. J. Maurer}
 \email{r.maurer@warwick.ac.uk}
 \affiliation{Department of Chemistry, University of Warwick, Gibbet Hill Road, CV4 7AL Coventry, UK}

\date{\today}

\begin{abstract}
	The emergence of machine learning methods in quantum chemistry provides new methods to revisit an old problem: Can the predictive accuracy of electronic structure calculations be decoupled from their numerical bottlenecks? Previous attempts to answer this question have, among other methods, given rise to semi-empirical quantum chemistry in minimal basis representation. We present an adaptation of the recently proposed SchNet for Orbitals (SchNOrb) deep convolutional neural network model [Nature Commun. 10, 5024 (2019)] for electronic wave functions in an optimised quasi-atomic minimal basis representation. For five organic molecules ranging from 5 to 13 heavy atoms, the model accurately predicts molecular orbital energies and wavefunctions and provides access to derived properties for chemical bonding analysis. Particularly for larger molecules, the model outperforms the original atomic-orbital-based SchNOrb method in terms of accuracy and scaling. We conclude by discussing the future potential of this approach in quantum chemical workflows.
\end{abstract}


\maketitle

\section{\label{introduction} Introduction  }
Machine learning (ML) methods are, by now, firmly established as important tools in the physical sciences~\cite{RevModPhys.91.045002,Butler2018} and, in particular, in computational molecular simulation and electronic structure theory.~\cite{Bartok2017,Dral2020} Supervised and unsupervised learning algorithms alike are being regularly used to tackle problems that first principles theory and classical molecular dynamics simulations cannot feasibly address on their own. This includes the construction of accurate interatomic potentials for molecules, materials,~\cite{Behler2007,Bartok2010,Behler2011} and gas-surface dynamics,~\cite{Behler2007b,Kolb2017,Zhang2019} the prediction of response and spectroscopic properties,~\cite{Grisafi2018,Christensen2019} and the prediction of molecular and materials properties across chemical compound space to support molecule and catalyst design.~\cite{Saravanan2017, Kitchin2018}

As ML methods start to significantly influence computational simulation workflows,~\cite{Peterson2016, Goh2017, Butler2018} the question arises how ML and big data concepts can be neatly integrated into first principles methods. This is particularly relevant to eventually overcome a major limitation of quantum chemical methods: High-quality predictions come at the price of high computational effort in solving integrals and constructing coupled integro-differential equations. The strength of ML methods lies in their ability to efficiently reuse ab-initio data within computationally efficient models, while being mostly physically naive. By utilizing ML-based approaches to replace the computational bottlenecks of quantum chemistry and electronic structure theory, the correlation between predictive power and computational cost can be weakened. Several approaches have recently been put forward that follow this idea in different ways. An obvious route in this direction is to use neural networks as basis representation to directly solve the many-body problem~\cite{Sugawara2001,Manzhos2009, Hermann2019} or to use ML to develop density functional approximations~\cite{Brockherde2017}. 

A different route is to make existing quantum chemical methods computationally more feasible. This idea is as old as quantum chemistry and electronic structure theory itself. This problem has been addressed, in the past, by neglecting or approximating difficult computations, i.e. replacing explicit integrals with precalculated and tabulated parameters. The most popular methods based on Neglect of Differential Diatomic Overlap (NDDO) such as AM1~\cite{Dewar1985} or the Density-Functional-based Tight-Binding~\cite{Elstner98,Koskinen2009} method construct  parametrizations for minimal basis representations of Hamiltonians that retain an explicit description of the electronic structure at vastly reduced computational cost. ML methods have been used in this context, with recent publications reporting the ML-based construction of semi-empirical methods such as OM2,~\cite{Dral2015} and DFTB~\cite{Li2018}, as well as the ML-based predictions of correlation energies based on Hartree-Fock molecular orbitals.~\cite{Welborn2018, Cheng2019}

In a similar vein, we have recently proposed a deep convolutional neural network called SchNet for Orbitals (SchNOrb),~\cite{Schutt2019} which predicts molecular wave functions and molecular orbital (MO) energies from Hartree-Fock (HF) and Density Functional Theory (DFT) for molecules as a function of their atomic configuration. This was achieved by representing the Hamiltonian in the same local atomic orbital basis that is used to create the quantum chemical training data. MO energies for systems such as ethanol, malondialdehyde (MDA) and uracil can be predicted within chemical accuracy (1 kcal/mol). By directly representing the wave function, the SchNOrb model provides access to many derived quantities, which can be represented as quantum mechanical expectation values. The predicted wave function can be used to initialise electronic structure calculations or to perform inverse design optimisation of electronic properties. One of the key limitations of this model for large-scale application lies in the use of the full basis representation in which the quantum chemical training data is provided. As suggested in the original publication, this can be remedied by preprocessing the data to generate effective optimised minimal basis representations. This offers the prospect of achieving high predictive power for large molecular systems at low computational cost by creating minimal basis Hamiltonian models that reproduce the results of DFT or HF. Furthermore, this represents an important step towards the integration of ML-based wave function representations into existing semi-empirical software frameworks.

In this paper, we use the SchNOrb deep learning framework to train a minimal basis (one atomic orbital basis function per atomic electronic eigenstate) representation of molecular wave functions that exactly reproduce the valence MO spectrum of organic molecules such as  aspirin and salicylic acid using the quasi-atomic minimal-basis-set orbitals, so-called QUAMBOs, put forward by Lu \emph{et al}.~\cite{Lu2004,Lu2005} We find that, when SchNOrb is trained on QUAMBO-projected training data, orbital energies and coefficients are predicted much more accurately and evaluated more efficiently than when trained on full-basis representations.


\section{\label{methods} Methods}

\subsection{\label{schorb} The SchNet for Orbitals (SchNOrb) deep learning model}

SchNOrb is a deep learning framework which models electronic structure based on a local atomic orbital basis representation of the wavefunction. The use of an atomic orbital basis to describe the wavefunction is common practice in quantum chemistry.
Molecular wavefunctions in wave-function-based methods or Kohn-Sham DFT are typically constructed from single particle states or molecular orbitals (MOs) $\ket{\psi_m}$, which are constructed by a linear combination of atomic orbitals $\ket{\psi_m}~=~\sum_i~c_{im} \ket{\phi_i}$ possessing different radial components and angular momenta.
By introducing such a basis, the time-independent eigenvalue equation of the electronic ground state can be cast in a simple matrix form
\begin{equation}
\Hr \mathbf{c}_m = \epsilon_m \Sr \mathbf{c}_m. \label{eq:TISE}
\end{equation}
The $\mathbf{c}_m$ is the atomic orbital coefficient vector associated with molecular orbital $m$ and the matrices $\Hr$ and $\Sr$ are the Hamiltonian operator $\hat{H}$ and orbital overlaps projected onto the local atomic basis:
\begin{align}\label{eq:H_and_S}
H_{ij} &= \braket{\phi_i | \hat{H} | \phi_j}\\
S_{ij} &= \braket{\phi_i | \phi_j}
\end{align}

Expressing the electronic problem in an atomic basis also offers several advantages from a machine learning perspective.
The direct prediction of orbital energies $\epsilon_m$ and wavefunction coefficients $\mathbf{c}_m$ is problematic, as both quantities are not smooth functions of the atomic positions. 
Moreover, the coefficient vectors are only determined up to an arbitrary phase rendering training of a model difficult.
The matrices $\Hr$ and $\Sr$ on the other hand are much better behaved while containing the same fundamental information as the orbital energies and coefficients, both of which can be recovered by solving the eigenvalue problem in Eq.~\ref{eq:TISE}.
As a consequence, SchNOrb was designed to operate in a local atomic picture with $\Hr$ and $\Sr$ as central quantities.

In the following, it will be explained how these matrices are modeled in detail using $\Hr$ as an example.
In SchNOrb, the Hamiltonian $\Hr$ is decomposed into different blocks associated with pairs of atoms:
\begin{equation}
\mathbf{H} = \left[\begin{array}{ccccc}
\mathbf{H}_{11} & \cdots & \mathbf{H}_{1j} &\cdots & \mathbf{H}_{1n}\\ 
\vdots  & \ddots & \vdots & & \vdots\\ 
\mathbf{H}_{i1}  & \cdots &  \mathbf{H}_{ij} & \cdots &  \mathbf{H}_{in}\\ 
\vdots & & \vdots & \ddots &\vdots \\ 
\mathbf{H}_{n1} & \cdots & \mathbf{H}_{nj} &  \cdots & \mathbf{H}_{nn}\\
\end{array} \right] 
\end{equation}
We further differentiate between onsite blocks $\mathbf{H}^\mathrm{on}_{i=j}$ which involve only a single atom $i$ and offsite blocks $\mathbf{H}^\mathrm{off}_{i \neq j}$ between two different atoms $i$ and $j$. 
Each of these blocks is represented by symmetry-adapted, pairwise feature vectors $\mathbf{\Omega}_{ij}$.
In order to capture the rotational symmetries in $\Hr$ up to angular moment $l$ arising from the different angular momenta of the atomic basis functions, SchNOrb constructs these features as polynomials of increasing order
\begin{equation}
\mathbf{\Omega}_{ij}^{(l)} = \prod^{l}_{\lambda=0} \mathbf{\omega}^{(\lambda)}_{ij} \quad \text{with } 0 \leq l \leq 2L.\label{eq:omega}
\end{equation}
The $\mathbf{\omega}^{(\lambda)}_{ij}$ are linear combinations of pairwise basis functions generated in each layer $\lambda$ of SchNOrb:
\begin{equation}
\mathbf{\omega}^{(\lambda)}_{ij} = \begin{cases}
\,\, \mathbf{p}_{ij}^{(\lambda)} \otimes \Id_{D} & \text{for } \lambda=0\\
\left[ \mathbf{p}_{ij}^{(\lambda)} \otimes \frac{\mathbf{r}_{ij}}{\|\mathbf{r}_{ij}\|} \right]  \mathbf{W}^{(\lambda)} & \text{for } \lambda>0\\
\end{cases}
\end{equation}
Here, the $\mathbf{p}_{ij}^{(\lambda)}$ are the coefficients of the linear combination predicted by SchNOrb and can be thought of as the basis function coefficients in analogy to conventional quantum chemical basis sets.
By using directional cosine vectors as Cartesian basis functions, the resulting features are rotationally equivariant for $\lambda > 0$ (and invariant for $\lambda=0$)
$\mathbf{W}^{(\lambda)}$ are trained parameters that allow for mixing between different functions.
By using this product form, SchNOrb can construct polynomials of various orders capturing the geometric transformations of the target Hamiltonian.


To do so, a suitable set of $\mathbf{p}_{ij}^{(\lambda)}$ capturing all relevant interactions in $\Hr$ needs to be constructed.
As a starting point, SchNet features $\{\mathbf{x}_i\}$ are constructed for each atom $i$ using the standard SchNet neural network architecture.~\cite{Schutt2017, Schutt2018} These features are rotationally invariant descriptors of the atomic environment.

Using a factorized tensor layer, the SchNet features between two atoms $i$ and $j$ are then combined into pairwise representations $\mathbf{h}_{ij}$:
\begin{equation}
\mathbf{h}_{ij}^{(\lambda)} = \sigma \left( \, \mathrm{lin}_2 \left[ \mathrm{lin}_{1} (\mathbf{x}_i^{(\lambda)}) \circ \mathrm{lin}_{1}(\mathbf{x}_j^{(\lambda)}) \circ \mathbf{W}_\text{filter}(r_{ij}) \right] \right). \label{eq:tensorfac}
\end{equation}
$\lambda$ denotes the current layer of SchNOrb and '$\mathrm{lin}$' denotes a linear neural network layer of the form
\begin{equation}
\mathrm{lin}(\mathbf{x}_i) = \mathbf{W}^\top \mathbf{x}_i + \mathbf{b},
\end{equation}
where $\mathbf{W}$ is a matrix of weights and $\mathbf{b}$ a vector of bias offsets, representing the trainable parameters of the network. Different subscripts denote the use of different sets of parameters.
$\sigma(x)$ is the shifted softplus activation function
\begin{equation}
\sigma(x) = \ln\left(\frac{1}{2}e^x + \frac{1}{2}  \right).
\end{equation}  
Finally, $\mathbf{W}_\text{filter}(r_{ij})$ is the radial filter function introduced in Ref.~\citenum{Schutt2018}, accounting for the distance dependence of the pairwise features. 
The radial filter also introduces a cosine based cutoff to spatially confine the interaction region.\cite{Schutt2019}
The factorized tensor layer operates in the same manner as introduced in Ref.~\citenum{schutt2017quantum}.
The features of both atoms are first projected to a factor space ($\mathrm{lin}_{1}: \mathbb{R}^{D} \mapsto \mathbb{R}^{F}$). There, both atoms interact and are weighted by $\mathbf{W}_\text{filter}(r_{ij})$. The result is then transformed back into feature space ($\mathrm{lin}_{2}: \mathbb{R}^{F} \mapsto \mathbb{R}^{D}$).

With the pairwise features $\mathbf{h}_{ij}^{(\lambda)}$ at hand, SchNOrb proceeds in a similar manner as the original SchNet architecture. The atomic features $\mathbf{x}_i^{(\lambda)}$ are refined over the course of several iterations, capturing patterns of increasing complexity:
\begin{equation}
\mathbf{x}_i^{(\lambda+1)} = \mathbf{x}_i^{(\lambda)} + \mathrm{mlp}_\mathrm{atom}\left(\sum_j \mathbf{h}_{ij}^{(\lambda)} \right),
\end{equation}
where $\mathrm{mlp}$ stands for a 2-layer feed forward network of the form 
\begin{equation}
\mathrm{mlp}\left(\mathbf{x}_i\right) = \mathrm{lin}_{a}\left( \sigma[\mathrm{lin}_{a}(\mathbf{x}_i)]\right).
\end{equation}

At the same time, the set of pairwise basis function coefficients $\mathbf{p}_{ij}^{(\lambda)}$ for the current interaction layer $\lambda$ are constructed as:
\begin{align}
\mathbf{p}_{ij}^{(\lambda)} &= \mathbf{p}_{ij}^{(\lambda, \mathrm{pair})} + \sum_{m \neq i} \mathbf{p}_{mj}^{(\lambda, \mathrm{env})} + \sum_{n \neq j} \mathbf{p}_{in}^{(\lambda, \mathrm{env})} \label{eq:coeff}\\
\mathbf{p}_{ij}^{(\lambda, \mathrm{pair})} &= \mathrm{mlp}_\mathrm{pair}\left(\mathbf{h}_{ij}^{(\lambda)}\right)\\
\mathbf{p}_{ij}^{(\lambda, \mathrm{env})} &= \mathrm{mlp}_\mathrm{env}\left(\mathbf{h}_{ij}^{(\lambda)}\right).
\end{align}
The first term on the right-hand side in Eq. \ref{eq:coeff} models the direct interaction between both atoms, while the remaining terms account for the influence of neighboring atoms.
A next set of $\mathbf{h}_{ij}^{(\lambda+1)}$ is then determined based on the updated features $\mathbf{x}_{i}^{(\lambda)}$ and Eq.~\ref{eq:tensorfac} etc.

After several iterations, the pairwise feature vectors $\mathbf{\Omega}_{ij}^{(l)}$ are constructed according to Eq.~\ref{eq:omega} and used to model the matrix blocks
\begin{align}
\mathbf{H}_{i}^\mathrm{on} &= \sum_l \sum_j \mathrm{linear}^l_\mathrm{on}\left( \mathbf{\Omega}_{ij}^{(l)}  \right) \\
\mathbf{H}_{ij}^\mathrm{off} &= \sum_l \mathrm{linear}^l_\mathrm{off}\left( \mathbf{\Omega}_{ij}^{(l)} \right).
\end{align}
The onsite and offsite blocks are then assembled into the predicted Hamiltonian, which is symmetrized in a final step.
\begin{equation}
\Hr = \frac{1}{2} \left( \Hr + \Hr^\top \right).
\end{equation}
The construction of $\Sr$ proceeds in an analogous fashion.

\subsection{\label{quambo} Quasi-Atomic Minimal Basis Orbitals (QUAMBOs)}

Electronic structure theory calculations provide more accurate results if wave functions and electron densities are expanded in larger and more complete basis sets. However, the computational cost of semi-local DFT and Hartree-Fock calculations nominally scales as $N^3$ and $N^4$ with the number of basis functions $N$. Semi-empirical methods and tight-binding approximations in chemistry have always tried to circumvent this issue by representing the electronic structure in a minimal or very  small  basis  set  of  local  atomic  orbitals,   which are  optimized  to  best  reproduce  the  results  of  full basis  calculations. Examples of such efforts include projecting onto a minimal basis of local atomic orbitals (AOs),~\cite{Agapito2016} maximally localized Wannier functions,~\cite{Marzari2012}  atomic orbitals in confinement potentials,~\cite{Eschrig1978, Smutna2020} and various orbital localization schemes such as quasi-atomic minimal-basis-set orbitals (QUAMBOs).~\cite{Lu2004,Lu2004b, Lu2005} 

In the context of creating a machine learning representation of electronic structure, transforming DFT data into a minimal representation of AOs offers important advantages: The projection onto a minimal basis reduces the number of matrix elements that need to be represented by a model of the Hamiltonian. As such, the training data is compressed into the smallest possible representation that still retains all relevant information. This has the additional benefit of reducing the scaling prefactor when diagonalizing the Hamiltonian. 

The QUAMBOs are a quasi-atomic basis representation designed with this condition in mind whilst possessing a few added benefits. For instance, the construction of QUAMBOs is completely general and independent of the full basis used in the electronic structure calculation. In the past, QUAMBOs have been constructed from local atomic orbitals~\cite{Lu2004b} and from Bloch waves~\cite{Lu2005}. Transforming a Hamiltonian from a full basis into a QUAMBO representation, retains a small subset of the molecular orbital space, typically the occupied MOs, while reducing the virtual orbital space, which does not affect the total energy determined by the self-consistent-field (SCF) equations used in HF and semi-local Kohn-Sham DFT.

To construct a set of $M$ QUAMBOs $\{\ket{A_j}\}$, we first define a subset of $M$ minimal basis AOs $\{\ket{A_j^0}\}$ of our full-basis AOs $\ket{\phi_i}$. This subset can be chosen arbitrarily, but here we choose the minimal set of atomic eigenstates that comprises the molecular system (e.g. H\textsubscript{2}O has seven minimal AOs, five associated with $O$ and one associated with each $H$ atom). We can define this minimal free-atom AO basis simply by projection from the $N$ occupied SCF MOs $\ket{\psi_n}$ and $V$ unoccupied SCF MOs $\ket{\psi_v}$, where $N+V$ equals the total number of MOs:
\begin{equation}
\ket{A_j^0} = \sum_n^N \ket{\psi_{n}}a_{nj}^0 + \sum_v^V \ket{\psi_{v}}a_{vj}^0
\end{equation}
with
\begin{equation}
a_{nj}^0 = \braket{\psi_{n}|A_j^0}
\end{equation}
and
\begin{equation}
a_{vj}^0 = \braket{\psi_{v}|A_j^0} .
\end{equation}
In the case where $\ket{A_j^0}$ are a subset of the full basis, $a_{nj}^0$ and $a_{vj}^0$ are simply a subset of the wave function coefficients $\mathbf{c}$ defined in eq.~\ref{eq:TISE}.

QUAMBOs are constructed similarly, but use a $P$-dimensional subspace of the $V$ unoccupied SCF MOs:
\begin{equation}\label{eq:quambo_expansion}
\ket{A_j} = \sum_n^N \ket{\psi_n}a_{nj} + \sum_p^P    \ket{\tilde{\psi}_p} b_{pj}
\end{equation}
The expansion in eq.~\ref{eq:quambo_expansion} ensures that the QUAMBOs correctly capture the occupied MO space by mixing occupied MOs with a small set of $P$ principal components of the virtual MO linear combinations $\ket{\tilde{\psi}_p}$ (`virtual valence states', VVS):
\begin{equation}
\ket{\tilde{\psi}_p} = \sum_v^V T_{pv} \ket{\psi_v}.
\end{equation}
The coefficients $T_{pv}$ are constructed from the virtual states by principal component analysis (\textit{vide infra}). The small set of $P$ VVS are chosen such that $M=N+P$ equals the number of minimal free-atom AOs. For example, in the case of H$_2$O, there are 5 occupied MOs ($N=5$) and 7 minimal basis AOs ($M=7$), therefore 2 VVS are needed ($P=2$).


As described in detail by Lu \textit{et al},~\cite{Lu2004} QUAMBOs are determined by a Lagrange minimisation of the mean square displacement between the  minimal AO basis $\ket{A_j^0}$ and the QUAMBO AOs $\ket{A_j}$ under the constraint of normalisation. This requires the QUAMBOs to remain as close as possible to AOs while still correctly reproducing the eigenspectrum of the full basis Hamiltonian. This yields following definition for the QUAMBOs:
\begin{equation}\label{eq:quambo}
\ket{A_j} = D^{-1/2}_j\left[ \sum_n^N \ket{\psi_n} a_{nj}^0 + \sum_p^P \ket{\tilde{\psi}_p} b_{pj} \right]
\end{equation}
where
\begin{equation}\label{eq:norm}
D_j = \sum_n |a_{nj}^0|^2 + \sum_p |b_{pj}|^2
\end{equation}
and
\begin{equation}
b_{pj} =  \braket{\tilde{\psi}_p|A_j^0} .
\end{equation}

In practice, we perform following steps to construct QUAMBOs and to transform Hamiltonian $\mathbf{H}$ and overlap matrix $\mathbf{S}$ into QUAMBO representation:
\begin{enumerate}
	\item We generate a set of orthonormal MOs by L\"owdin orthogonalisation as a starting point for the transformation:
	\begin{equation}
	\mathbf{\Psi^O} = \mathbf{\Psi} \mathbf{S}^{-1/2} 
	\end{equation}
	Note that this yields slightly different QUAMBOs to when starting from canonical MOs followed by subsequent intra-atomic orthogonalisation as originally proposed by Lu \emph{et al}.~\cite{Lu2004}
	\item Select a minimal basis set definition $\ket{A_j^0}$ and separate orthogonalised MOs into `occupied' (N) and virtual space (V) in terms of $N$ projection coefficients $a_{nj}^0$ and $V$ coefficients $a_{vj}^0$ for the occupied and virtual space, respectively.\footnote{Note that the occupied space can also contain nominally unoccupied orbitals as long as $N<M$.}
	\item Use the virtual space projection coefficients $a_{vj}^0$ to construct a matrix $B_{vw}=\sum_j a_{vj}^0a_{jw}^{*0}$ of dimension $V\times V$.
	\item Select the $P$ largest eigenvalues of $\mathbf{B}$ to create a $P\times V$ matrix $\mathbf{T}$ from the eigenvector columns of $\mathbf{B}$ associated with those eigenvalues.
	\item Use $\mathbf{T}$ to create the set of VVS:
	\begin{equation}
	\tilde{\mathbf{\Psi}}_p = \textbf{T}_p \mathbf{\Psi}_v
	\end{equation}
	\item Use eq.~\ref{eq:quambo} to construct the QUAMBO states. We arrive at a matrix $\mathbf{A}$ with dimension $M\times(N+V)$, defining the $M$ QUAMBOs in terms of the original AO basis set with dimension $(N+V)$. The created QUAMBOs are nonorthogonal.
	\item We transform $\mathbf{H}$ and $\mathbf{S}$ into QUAMBO representation via:
	\begin{equation}
	\textbf{H}_{Q} = \textbf{A}^T \textbf{H} \textbf{A}
	\end{equation}
	and
	\begin{equation}
	\textbf{S}_{Q} = \textbf{A}^T \textbf{A} .
	\end{equation}
\end{enumerate}
A further L{\"o}wdin orthogonalisation yields an orthogonal QUAMBO Hamiltonian, if desired:
\begin{equation}
\textbf{H}_{Q}^{\text{Ortho}} = \textbf{S}_{Q}^{-\frac{1}{2}} \textbf{H}_{Q} \textbf{S}_{Q}^{-\frac{1}{2}} .
\end{equation}

\begin{figure}
	\includegraphics[width=3.3in]{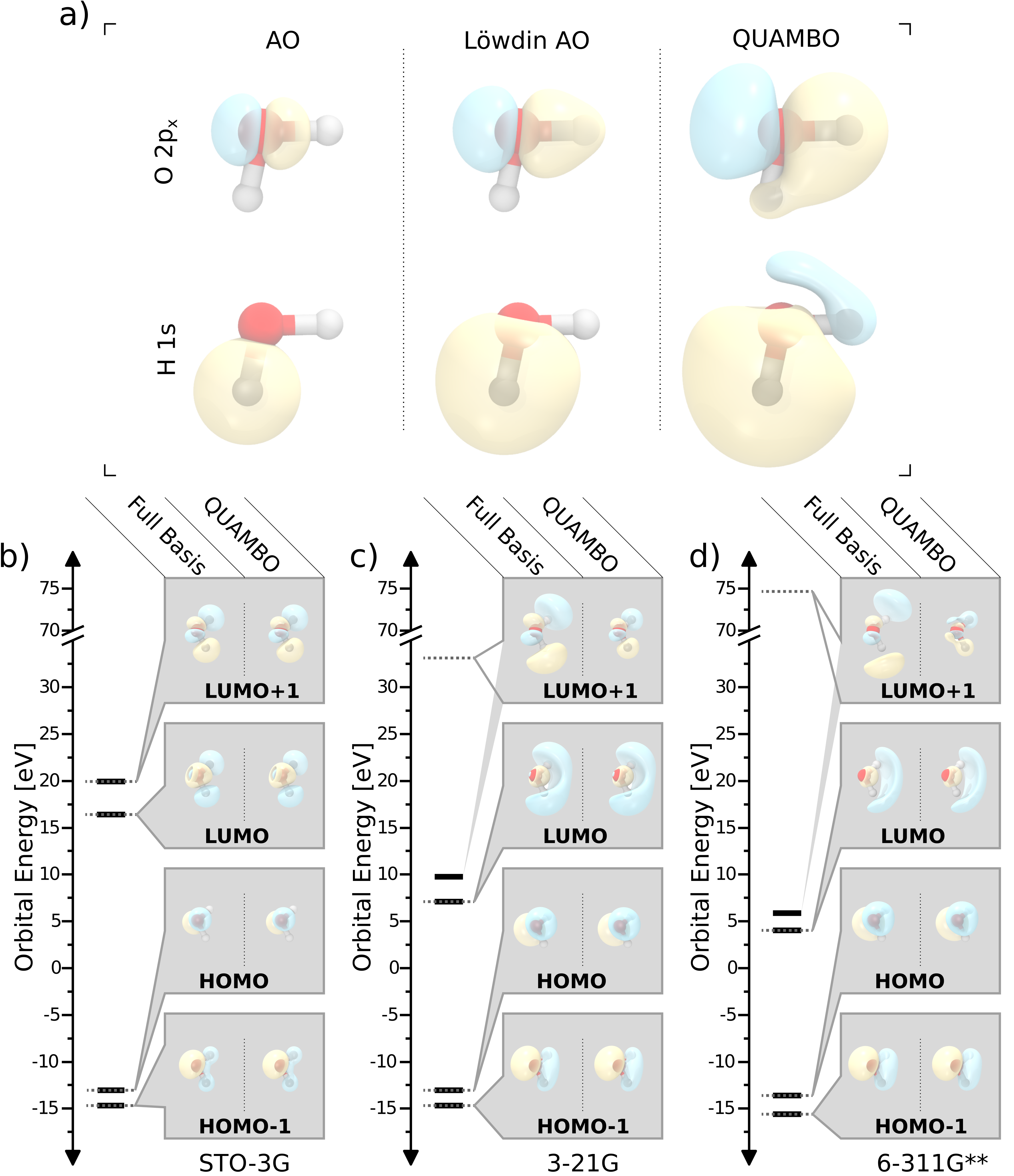}
	\caption{(a) Depiction of different basis states for Oxygen and Hydrogen. From left to right: full basis atomic orbital (AO), L\"odwin orthogonalised AO, QUAMBO quasi-atomic orbital. (b-d) MO energies and orbital representations for eigenstates of full basis $\mathbf{H}$ and $\mathbf{S}$ (`Full Basis') and for eigenstates of QUAMBO-transformed $\mathbf{H}_Q$ and $\mathbf{S}_Q$ (QUAMBO). Figure shows results for three different full basis cases: (b) STO-3G minimal basis, (c) 3-21G double-zeta basis, and (d) 6-311G** triple-zeta basis. Full basis energy levels are depicted as thick lines, QUAMBO-transformed MO energy levels are shown as dotted lines. Insets show 2 highest occupied and 2 lowest unoccupied orbitals (HOMO-1, HOMO, LUMO, LUMO+1).}
	\label{fig:h2o-quambos}
\end{figure}

Figure~\ref{fig:h2o-quambos} compares MO energy levels and orbital representations in full basis and after QUAMBO transformation in different Gaussian-type basis sets. In all three cases, we select a  subset of the basis functions as the minimal basis $\ket{A^0_j}$. The underlying calculations have been performed with Orca~\cite{Neese2012} and a PBE functional.~\cite{Perdew1996} In the QUAMBO transformation, we define the `occupied' subspace to contain all occupied orbitals and the lowest unoccupied MO (LUMO), which means $N=6$ and $P=1$.  The STO-3G basis is a minimal basis with 7 AOs. Therefore there is no dimensionality reduction in this case and AOs and QUAMBOs are identical. The 3-21G and 6-311G** basis sets contain 13 and 30 AOs, respectively, which are transformed into 7 QUAMBOs. In both cases, the QUAMBO projection correctly retains the MO energies and MO shapes for the 6 lowest MOs, but, as expected, does not conserve the energy or shape of the LUMO+1. This MO does not contribute to the total energy or any other ground-state property of the molecule.
As such, the QUAMBO transformation provides for an efficient Hamiltonian dimensionality reduction without loss of information or accuracy, which we will use to preprocess our training data for the SchNOrb model.

\subsection{\label{data-augmentation} Data Augmentation}

The features $\Omega_{ij}^{(l)}$ in SchNOrb are constructed in a general manner, allowing them to capture arbitrary rotational dependencies in the modeled quantities.
Since no \emph{a priori} constraints are imposed, the correct rotational symmetry of the orbitals is learned in a purely data-driven fashion.
As a consequence, sufficient rotations of a molecule need to be present during the training procedure in order to yield meaningful models of molecular wavefunctions.
We achieve this, by sampling random molecular rotations for each data point during the training procedure.
Since the entries of the Hamiltonian and overlap matrices exhibit specific symmetries depending on the angular momenta of the orbitals involved, they can be transformed using Wigner $\mathcal{D}$ rotation matrices.
After sampling a random rotor $R$, the following transformations can be applied
\begin{align}
\tilde{\mathbf{R}}_{i} &= {\mathcal{D}}^{(1)}(R) \mathbf{R}_i \\
\tilde{\mathbf{F}}_{i} &= {\mathcal{D}}^{(1)}(R) \mathbf{F}_i \\
\tilde{\mathbf{H}}_{\mu \nu} &= {\mathcal{D}}^{(l_\mu)}(R) \mathbf{H}_{\mu \nu}  {\mathcal{D}}^{(l_\nu)}(R) \\
\tilde{\mathbf{S}}_{\mu \nu} &= {\mathcal{D}}^{(l_\mu)}(R) \mathbf{S}_{\mu \nu}  {\mathcal{D}}^{(l_\nu)}(R)
\end{align}
in order to generate the rotated positions $\tilde{\mathbf{R}}_i$, atomic forces $\tilde{\mathbf{F}}_i$, Hamiltonian matrices $\tilde{\mathbf{H}}$, and overlap $\tilde{\mathbf{S}}$:
$\mathbf{H}_{\mu \nu}$ and $\mathbf{S}_{\mu \nu}$ correspond to blocks in the Hamiltonian and overlap matrices between atomic basis functions with the angular momenta $l=\mu$ and $l=\nu$.
$\mathcal{D}^{(l_\mu)}$ is the Wigner $\mathcal{D}$ matrix for angular momentum $l=\mu$, while $\mathcal{D}^{(1)}$ is the rotation matrix in 3D space.

In case of QUAMBO Hamiltonians $\mathbf{H}_Q$ and overlap matrices $\mathbf{S}_Q$, the individual basis functions are no longer pure in angular momentum.
As such, it is not possible to construct a Wigner $\mathcal{D}$ matrix to rotate the QUAMBO matrices directly.
Instead, we first rotate the original Hamiltonian and overlap matrix with the procedure described above.
Afterwards, we apply the QUAMBO projection to the rotated matrices in order to obtain the rotated matrices $\tilde{\mathbf{H}}_Q$ and $\tilde{\mathbf{S}}_Q$ in QUAMBO representation.

\subsection{\label{computational_details} Computational Details}

The reference data for malondialdehyde and uracil were taken from a previous study\cite{Schutt2019}.
For the remaining molecules, new reference data were generated.
To this end, 30,000 reference configurations were sampled at random from the MD17 datasets for the molecules asparagine, salicylic acid and aspirin\cite{chmiela2017machine}.
In order to be consistent with the previous study, reference computations for the selected structures were carried out with the ORCA quantum chemistry code at the PBE/def2-SVP level of theory\cite{Neese2012,Perdew1996, Weigend2005PCCP}.
Integration grid levels of 4 and 5 were employed during SCF iterations and the final computation of properties, respectively and SCF convergence criteria were set to 'very tight'.

The proton transfer in malondialdehyde shown in Figure 6 was sampled via a molecular dynamics simulation using an integration time step of 0.5~fs for a total of 50~ps.
The simulation temperature was kept at 300~K using a Langevin thermostat with a time constant of 100~fs.

QUAMBO preprocessing and model training was implemented by adapting an existing SchNOrb repository based on SchNetPack\cite{schuett2018schnetpack}.
In all experiments -- full basis and QUAMBO -- SchNOrb architectures were constructed using three SchNet interaction blocks and five SchNOrb interactions.
SchNOrb feature vectors were set to a length of 1000 and a cosine cutoff of 10~\AA\ was applied to localize interactions (see above).
For training, the reference data sets were split into training, validation and test sets. 
Training was performed on the training sets, using the validation sets for monitoring the training progress.
The test sets were withheld from training and used for evaluating the performance of the resulting models.
Data splits for the individual molecule data sets can be found in Tab.~S1 in the SI, alongside mini batch sizes.
All models were trained on two GPUs using the ADAM optimizer\cite{kingma2014adam} and combined loss described in Ref.~\cite{Schutt2019} with a energy tradeoff of $10^{-3}$.
The initial learning rate of $10^{-4}$ was reduced using a decay factor of 0.8 after 50 epochs without improvement of the validation loss. The training was stopped when the learning rate dropped below a threshold of $5\times10^{-6}$.
After training, the best performing models were selected based on the validation error. We report errors as mean absolute errors (MAE). For Hamiltonian and overlap matrices, MAE is calculated as the average over all matrix elements and all sample errors.

\section{\label{results} Results and Discussion}

\subsection{\label{results-schnorb} SchNOrb Prediction Performance}

\begin{figure*}
	\centering
	\includegraphics[width=\textwidth]{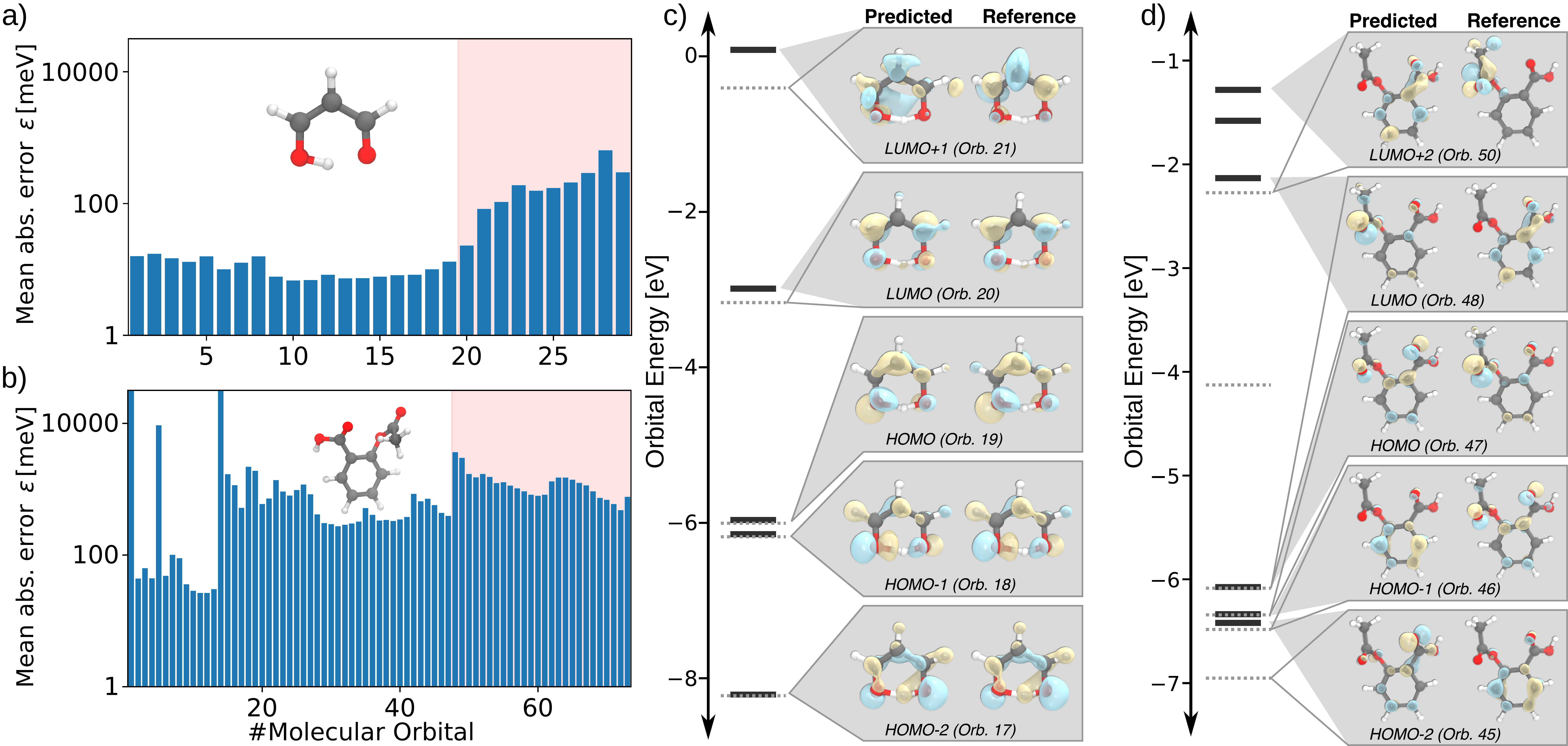}
	\caption{Accuracy of SchNOrb predictions for orbital energies $\epsilon$ and shapes using the full basis. a) Mean absolute errors of the individual molecular orbital energy levels in malondialdehyde. A red background indicates unoccupied molecular orbitals. b) Errors for aspirin. In both cases, the higher unoccupied states are omitted for clarity. c) Predicted and reference orbital energies and shapes for a single malondialdehyde configuration. Solid arrows indicate the assignment between reference orbitals and energies, while the outlined arrows indicate the assignment for the predicted quantities. d) Same as c) but with aspirin as prediction target.}
	\label{fig:results_schnorb_performance}
\end{figure*}

It was found in a previous study, that the SchNOrb network exhibits excellent performance in the prediction of the Hamiltonians and associated wavefunctions for small to medium sized organic molecules.~\cite{Schutt2019}
An example is depicted in Fig.~\ref{fig:results_schnorb_performance}a, which shows the mean absolute errors associated with the eigenvalues predicted by a SchNOrb model for the malondialdehyde (MDA) molecule using a test set consisting of approx. $1500$ configurations not used during training.
Whereas the SchNOrb architecture yields excellent predictions of occupied molecular orbitals and their corresponding energies, a degradation in prediction accuracy is found for virtual orbitals.
The latter can be rationalised due to the nature of the SCF problem, where virtual orbitals are not strictly bound by the Kohn-Sham equations and do not enter the ground-state electron density. Therefore, a larger variability of virtual orbitals as a function of configuration space is to be expected.
Fig.~\ref{fig:results_schnorb_performance}c shows the highest occupied and lowest unoccupied molecular orbitals and their energies for a sample configuration of MDA as computed with SchNOrb and the electronic structure reference.
For all occupied orbitals (HOMO-2 to HOMO) orbital shapes as well as energies predicted by the ML model agree closely with the reference.
For higher lying virtual states ($>$LUMO), the degradation described above is observed for the energy levels, as well as the shape of the associated orbital coefficients.
Nevertheless, SchNOrb still manages to reconstruct the LUMO orbital with reasonable accuracy. 
We have previously also demonstrated that derived properties such as atomic net charges and correlation energies can be predicted successfully based on SchNOrb model wavefunctions.

Yet, despite their overall reliability for medium sized molecules, SchNOrb models based on a similar number of training data points begin to exhibit problematic behavior when used to model systems with a larger number of electrons ($>$60).
This can be seen in Fig.~\ref{fig:results_schnorb_performance}b, where the test set orbital energy error is shown for the aspirin molecule. 
On the whole, the observed errors are larger by one order of magnitude, which could in principle be attributed to the increase in the dimensionality of the Hamiltonian and overlap matrices introduced by the additional electrons.
In addition to the systematic decrease in accuracy, very large errors ($>1$~eV) can be observed for certain energy levels.
The reason for these patterns in the eigenvalue spectrum become apparent, when studying the molecular orbitals close to the HOMO and LUMO (Fig.~\ref{fig:results_schnorb_performance}d) for a single aspirin conformation.
At a first glance, the SchNOrb predictions fail to yield accurate orbital energies and shapes, with the SchNOrb levels shifted significantly compared to the electronic structure reference.
Upon closer inspection of the molecular orbitals, it is found that this pathological behavior is primarily due to the insertion of orbitals by SchNOrb, which have no counterpart in the electronic structure, e.g. orbital 45 close to $-7$~eV and orbital 49 close to $-2$~eV (the latter not visualized in Fig.~\ref{fig:results_schnorb_performance}d).
This fundamentally changes the assignment of energy levels, leading to the poor model performance.
When comparing reference and predicted orbitals similar in energy instead, a closer correspondence between both methods is observed.
The SchNOrb HOMO-1 for example is the same as the reference HOMO-2, which also lies closer in energy than the SchNOrb counterpart assigned when only considering the order of states.

This unstable behavior is due to several reasons.
As stated above, the additional electronic degrees of freedom lead to larger Hamiltonian and overlap matrices (growing quadratically in size with the number of basis functions) which have to be modeled by the SchNOrb architecture.
Due to the increased size of the matrices, model errors can be distributed in many different ways.
As a consequence, it is possible to obtain good fits for the matrices, while the solution of the eigenvalue problem required for the orbital energies and coefficients becomes ill-conditioned due to error accumulation, leading e.g. to the intruding states observed for aspirin.
This behavior is further exasperated by the inherent flexibility of SchNOrb.
This flexibility makes it possible to learn the complicated rotational symmetries of Hamiltonian elements in a manner completely agnostic to the angular momentum properties of the atomic orbitals.
However, it also requires the introduction of the data augmentation procedure described above to properly learn the relevant transformations.
This works well for smaller molecules, but the sampling becomes less effective for systems with more basis functions, due to the steep increase of additional degrees of freedom which would need to be sampled during training. As will be shown in the next section, this problem can be mitigated by preprocessing the training data with a transformation that limits the Hamiltonian and overlap matrix dimensions to a minimal basis representation.

\subsection{\label{results-quambo} SchNOrb Prediction Performance in QUAMBO representation}

\begin{figure}
	\centering
	\includegraphics{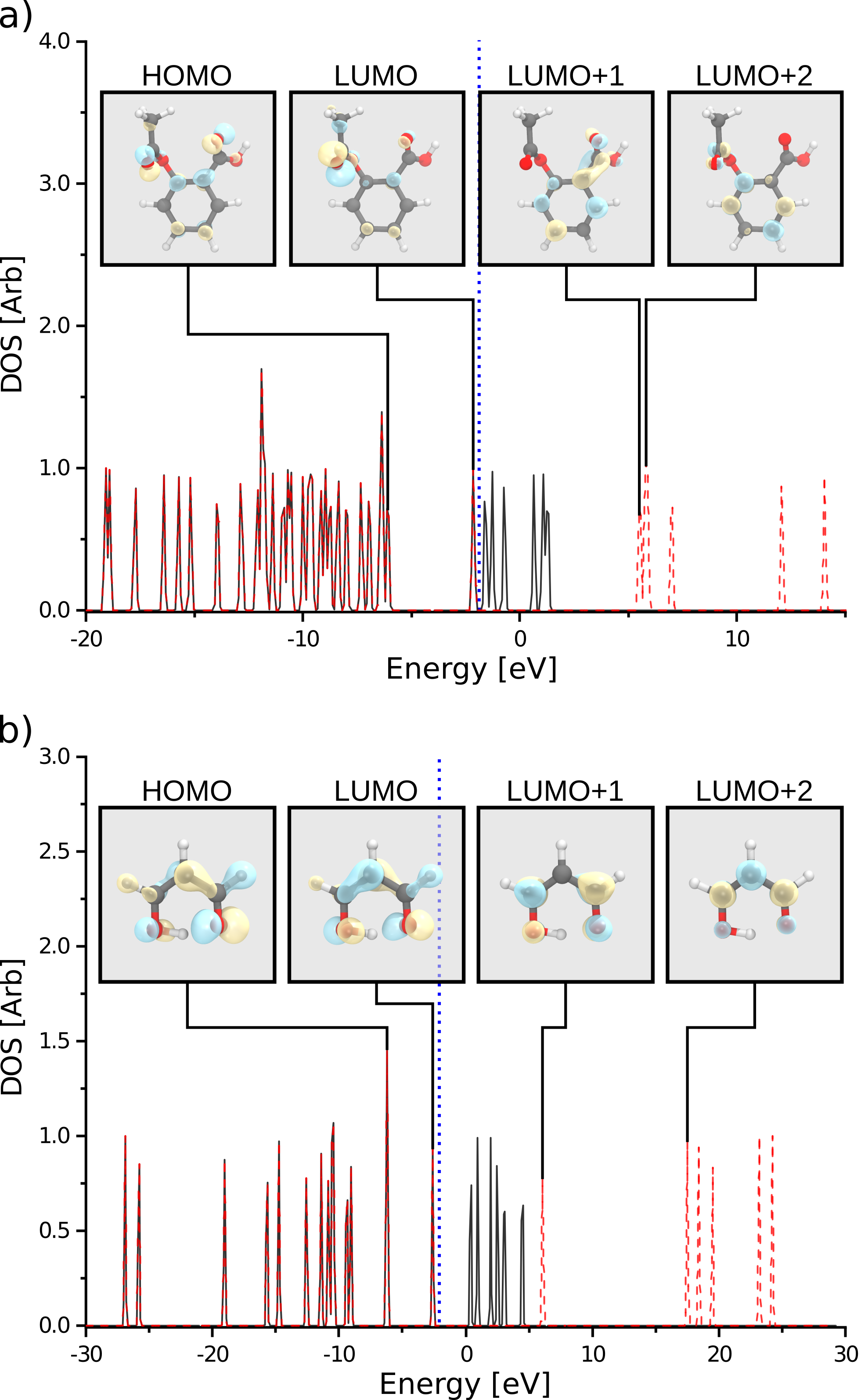}
	\caption{Full basis (black) and QUAMBO (red) densities of states for a) aspirin and b) MDA with insets visualising the HOMO-1, HOMO, LUMO, and LUMO+1 MOs after transformation; lines connect each inset to its associated QUAMBO eigenvalue. A dashed blue line is used to indicate the upper bounds of the conserved (`occupied') subspace.}
	\label{fig:results_quambo_visualization}
\end{figure}

We now compress the training data into a QUAMBO minimal basis representation by transforming the Hamiltonian and overlap matrices in full-basis AO representation. In this work, all QUAMBO transformations are performed with the LUMO included in the conserved (`occupied') subspace $\ket{\psi_n}$. We therefore expect that all occupied MO eigenvalues and the LUMO eigenvalue are exactly reproduced by the QUAMBO transformation. Figure~\ref{fig:results_quambo_visualization} shows the eigenvalue spectrum for an example configuration of aspirin and MDA (panels a and b, respectively) before and after QUAMBO transformation. We find that, indeed, the occupied eigenvalue space and the LUMO energy are exactly retained in the QUAMBO representation in both cases. As shown in the insets, the spatial distribution of the conserved MOs is also retained. The same is not true for the virtual space, where both orbital energies and shapes are distinct from the original representation with the QUAMBO virtual MO energies shifted considerably to higher energy. This is of no great consequence as these MOs do not contribute to molecular ground state properties.

The QUAMBO transformation achieves a significant reduction in matrix size without loss of information. For example, the dimension of the $\mathbf{H}$ and $\mathbf{S}$ matrix for aspirin and MDA in the original def2-SVP basis is 222$\times$222 and 90$\times$90, respectively. In minimal basis they become 73$\times$73 and 29$\times$29, respectively. This dimensionality reduction is expected to reduce compounding errors during eigenvalue decomposition, which in turn improves the quality and stability of orbital energy and coefficient predictions. 

\begin{figure*}
	\centering
	\includegraphics[width=\textwidth]{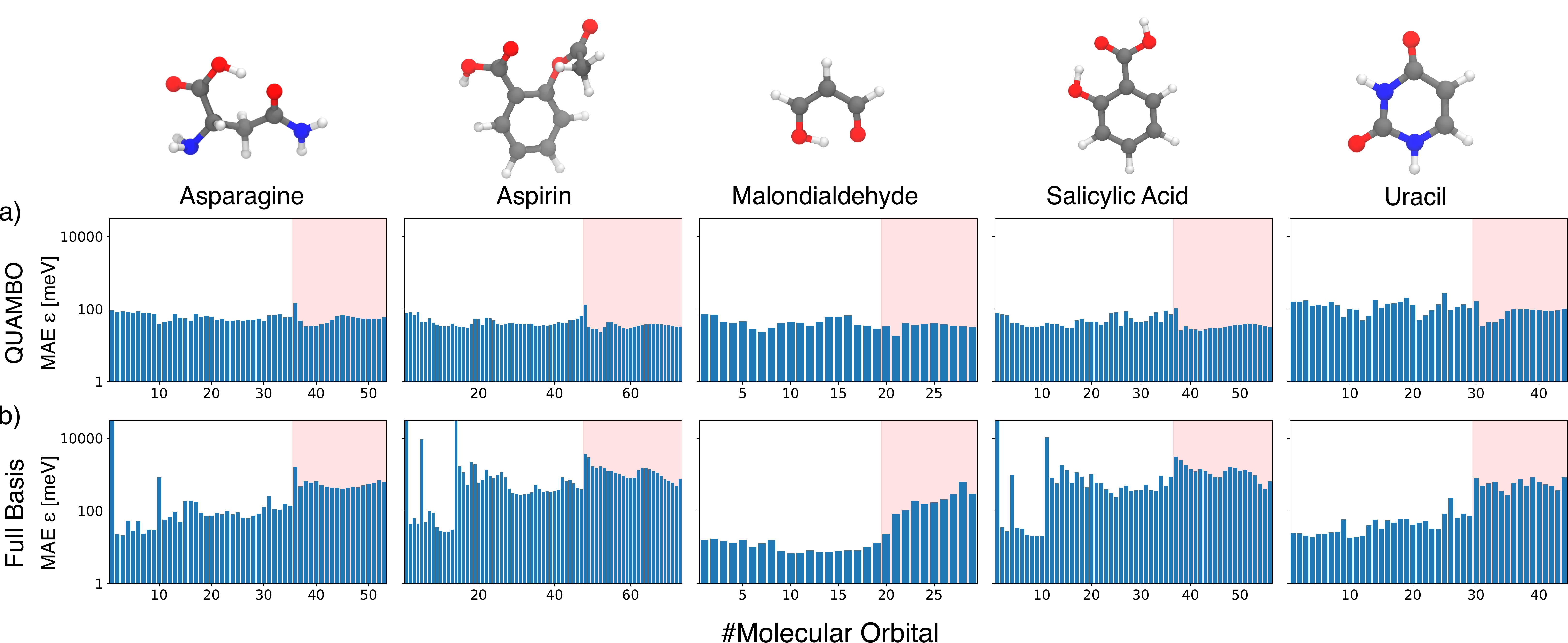}
	\caption{Comparison of the mean absolute test errors (MAE) associated with the molecular orbital energies $\epsilon$ predicted by SchNOrb models a) using the QUAMBO pre-processing step and b) the full basis set. In case of the latter, the same number of energy levels as obtained for the QUAMBO basis were included in the plot for reasons of clarity. A red background indicates unoccupied orbital space.}
	\label{fig:results_quambo_schnorb_performance1}
\end{figure*}

As can be seen in Fig.~\ref{fig:results_quambo_schnorb_performance1}, combining SchNOrb with a QUAMBO transformation (Fig.~\ref{fig:results_quambo_schnorb_performance1}a) does indeed lead to an overall improvement in the predicted molecular orbital energies when compared to a model trained in the full basis representation (Fig.~\ref{fig:results_quambo_schnorb_performance1}b).
The benefits of the QUAMBO pre-processing step are particularly pronounced for the larger molecules (asparagine, aspirin, salicylic acid).
There, the use of QUAMBOs completely eliminates the instabilities in the eigenvalue decomposition.
Erroneous spikes due to an unphysical reordering of orbital levels (observed e.g. for orbital 1) present in the full basis model are absent in the QUAMBO based SchNOrb predictions.
At the same time, the overall error for the remaining orbital energies is reduced by approximately one order of magnitude, demonstrating the utility of this approach.

Smaller molecules, such as uracil and malondialdehyde, appear to profit less from introducing QUAMBO transformed matrices.
For the former, using QUAMBOs leads to an increase in errors for the orbitals in occupied space.
There are two potential reasons for this effect.
First, the QUAMBO overlap matrices are significantly more complex than their conventional counterparts.
Due to the QUAMBO projection, the QUAMBO orbitals are no longer pure atomic states compared to the original overlap matrix.
This means, that the surrounding environment now influences the matrix entry between two basis functions, which was not the case in the full AO basis. 
As a result, the complexity to model the QUAMBO overlap matrix is similar to the Hamiltonian, leading to an overall increase in the prediction errors (see Tab.~\ref{tab:results}), which in turn influences the error associated with the predicted orbital energies.
To study this effect, we have repeated the analysis in Fig.~\ref{fig:results_quambo_schnorb_performance1} but this time using a combination of predicted Hamiltonians and reference overlap matrices and \textit{vice versa} for the computation of the predicted orbital energies.
The results of this experiment are collected in Fig.~S4 of the SI.
We find that the prediction errors associated with the overlap matrix primarily affect the lower eigenvalues, leading to the higher prediction errors observed in these regions compared to the original SchNOrb models.
A second contribution is related to the number of molecular degrees of freedom and the matrix size in the full basis.
Compared to the more compressed QUAMBO basis, the use of a full basis results in larger Hamiltonian and overlap matrices.
This leads to an increase in the overall number of model parameters, which in turn offers more flexibility for error redistribution during training.
As a consequence, the full basis model can achieve lower prediction errors in cases, where the molecule is still small enough for the data augmentation procedure to sample the configuration space of Hamiltonian and overlap matrices successfully.

Interestingly, the QUAMBO transformation improves predictions not only for occupied, but also for virtual orbital space (red background), even for the small molecules.
This indicates, that the QUAMBO virtual orbital coefficients vary less strongly as a function of configuration. This is likely a side product of mapping the effects of high angular momentum functions into an optimised effective minimal basis. The maximum prediction errors for all molecules are consistent and in a similar range as the mean absolute errors (see Fig~S1 in the SI).

\begin{table*}
	\centering
	\begin{tabular}{lrrrrrrrrrrrrrr}
		\toprule
		\multirow{2}{*}{Dataset} & \multicolumn{2}{c}{H [eV]} && \multicolumn{2}{c}{S} && \multicolumn{2}{c}{$\epsilon$ [eV]} && \multicolumn{2}{c}{Gap [eV]}  && \multicolumn{2}{c}{$\psi$} \\
		\cmidrule{2-3} \cmidrule{5-6}  \cmidrule{8-9}  \cmidrule{11-12} \cmidrule{14-15} 
		& full  & QUAMBO    && full  & QUAMBO   && full  & QUAMBO  && full & QUAMBO  && full & QUAMBO \\
		\midrule
		Asparagine      & 0.006838 &  0.016379 && 0.000076 & 0.000458 &&   1.8671 &  0.0629 && 1.4815  & 0.1491  && 0.84    & 0.96 \\
		Aspirin         & 0.005267 &  0.009984 && 0.000050 & 0.000281 &&  77.5683 &  0.0448 && 3.2554  & 0.1148  && 0.38    & 0.94 \\
		Malondialdehyde & 0.005153 &  0.009474 && 0.000067 & 0.000213 &&   0.0109 &  0.0449 && 0.0257  & 0.0422  && 0.99    & 0.99 \\
		Salicylic Acid  & 0.005353 &  0.010172 && 0.000060 & 0.000264 &&   4.4926 &  0.0507 && 2.2547  & 0.0792  && 0.46    & 0.96 \\
		Uracil          & 0.006238 &  0.018928 && 0.000082 & 0.000523 &&   0.0479 &  0.1261 && 0.7403  & 0.1206  && 0.90    & 0.96 \\
		\bottomrule
	\end{tabular}
	\caption{Mean absolute test errors for SchNOrb models using the full basis (full) and the QUAMBO transformed basis (QUAMBO) for the Hamiltonians $H$, overlap matrices $S$, orbital energies $\epsilon$, HOMO-LUMO gap (Gap) and mean unsigned Cosine similarity of the orbital coefficients $\psi$. The errors associated with the orbital energies and coefficients are computed using the occupied orbitals only.}
	\label{tab:results}
\end{table*}
Tab.~\ref{tab:results} summarises test errors of SchNOrb models trained on full basis sets and their respective QUAMBO representations for the Hamiltonians $H$, overlap matrices $S$, orbital energies $\epsilon$, HOMO-LUMO gap and orbital coefficients.
The QUAMBO projection offers significant advantages for quantities that rely on solving the eigenvalue equation (Eq.~\ref{eq:TISE}), namely the orbital energies, HOMO-LUMO gap and orbital coefficients.
Once again, the performance gain is most significant for the large molecules and aspirin in particular, where the mean absolute error on orbital energy predictions is reduced from almost 78~eV to 0.0448~eV.
For these systems, predictions of the HOMO-LUMO gap benefit in a similar manner and reductions in error of up to a factor of 30 can be observed compared to the full basis models.
On the whole, we also observe an improvement in the orbital coefficient accuracy (by a factor of 2 for aspirin and salicylic acid), suggesting that the QUAMBO pre-processing step yields significantly more accurate wavefunction representations.

While the derived quantities profit from the QUAMBO transformation in general, the errors associated with the directly predicted quantities (Hamiltonian and overlap matrix) are larger in the QUAMBO representation (Tab.~\ref{tab:results}).
In case of the Hamiltonians, different effects contribute. One of them is the reduced number of ways the errors can be distributed during training due to the smaller matrix size compared to the full basis mentioned previously. In addition, the different matrix sizes affect the error statistics, which makes an analysis based on a single averaged mean absolute error for the whole matrix more difficult.
\begin{figure*}
	\centering
	\includegraphics[width=\textwidth]{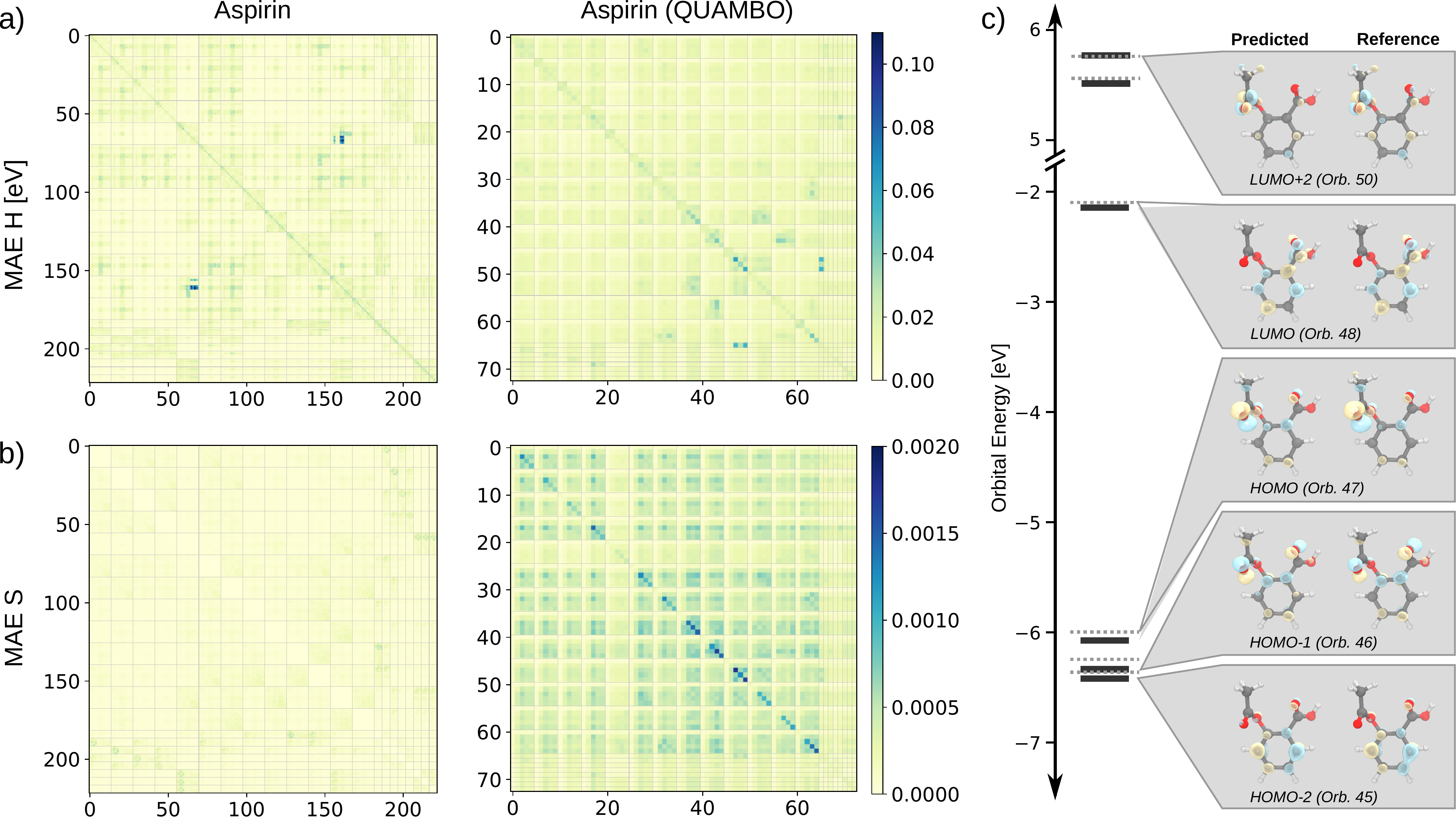}
	\caption{Analysis of prediction errors and orbital assignment in aspirin. a) Mean absolute test error for the individual elements of the full (left) and QUAMBO transformed Hamiltonian (right). Grey lines indicate the basis function blocks associated with individual atoms. b) Same as a) but for the overlap matrices. c) Predicted and reference orbital energies and shapes for a single aspirin configuration in the QUAMBO basis. Solid arrows indicate the assignment between reference orbitals and energies, while the outlined arrows indicate the assignment for the predicted quantities.}
	\label{fig:results_quambo_schnorb_performance2}
\end{figure*}
This effect is shown for aspirin in Fig.~\ref{fig:results_quambo_schnorb_performance2}a where the mean absolute errors associated with the individual Hamiltonian matrix elements are compared between full basis and QUAMBO based SchNOrb predictions.
The overall scale of the errors is the same, with lower maximal errors associated with the QUAMBO Hamiltonian.
However, many entries in the full Hamiltonian are close to zero and contribute little to the overall error.
Upon QUAMBO transformation, this sparsity is reduced due to the compression of information and the increased complexity of the underlying QUAMBO basis functions compared to the full basis AOs.
As a consequence, when calculating the mean absolute error by averaging over all matrix elements, the larger full basis matrix benefits more from sparsity than its less sparse QUAMBO counterpart. 
This observation also holds for the overlap matrix, with one additional effect to consider.
As stated previously, the QUAMBO orbitals are no longer pure atomic states compared to the original overlap matrix.
This increased complexity associated with modeling the QUAMBO overlap matrix leads to the overall increase in error of one order of magnitude.
The more complex structures in the QUAMBO overlap can be observed in Fig.~\ref{fig:results_quambo_schnorb_performance2}b, where the mean absolute errors for individual overlap matrix elements are visualized for aspirin. Similar effects are found for the other molecules. The associated errors in the elements of the Hamiltonians and overlap matrices are provided in Fig.~S2 of the SI.
SchNorb can also predict the total energy simultaneously to H and S and we analyse how the QUAMBO transformation affects this. The test errors predicted for the total energy can be found in Tab.~S3 of the SI.
The total energy prediction is only weakly coupled to the prediction of Hamiltonian and overlap matrices,  which are the main focus of this study.
Consequently, the loss function trade-off governing this coupling was not optimized for the present application, leading to higher errors than in the original SchNOrb manuscript.
However, even in this case, all models still achieve high prediction accuracy and an error of similar magnitude as in the full basis model can be recovered by choosing an appropriate value for the trade-off, as was demonstrated for a retrained QUAMBO-SchNOrb model of malondialdehyde (see Tab.~S4).
Interestingly, although only weakly coupled, the prediction of orbital energies appears to profit from a stronger weighting of the total energy loss during training.

Fig.~\ref{fig:results_quambo_schnorb_performance2}c compares the predicted and reference orbital energies and shape of aspirin for the same MOs as shown in Fig.~\ref{fig:results_schnorb_performance}d, but this time using the QUAMBO transformed basis. 
The advantage of using the reduced basis becomes immediately apparent, as it eliminates the presence of erroneous states and reproduces the correct ordering of orbitals.
As stated above, the QUAMBO projection also improves the SchNOrb prediction for the orbitals in virtual space.
This is advantageous if e.g. properties such as the HOMO-LUMO gap need to be studied, provided the virtual orbitals of interest are part of the conserved subspace in the QUAMBO projection.

\begin{figure}
	\centering
	\includegraphics[width=\columnwidth]{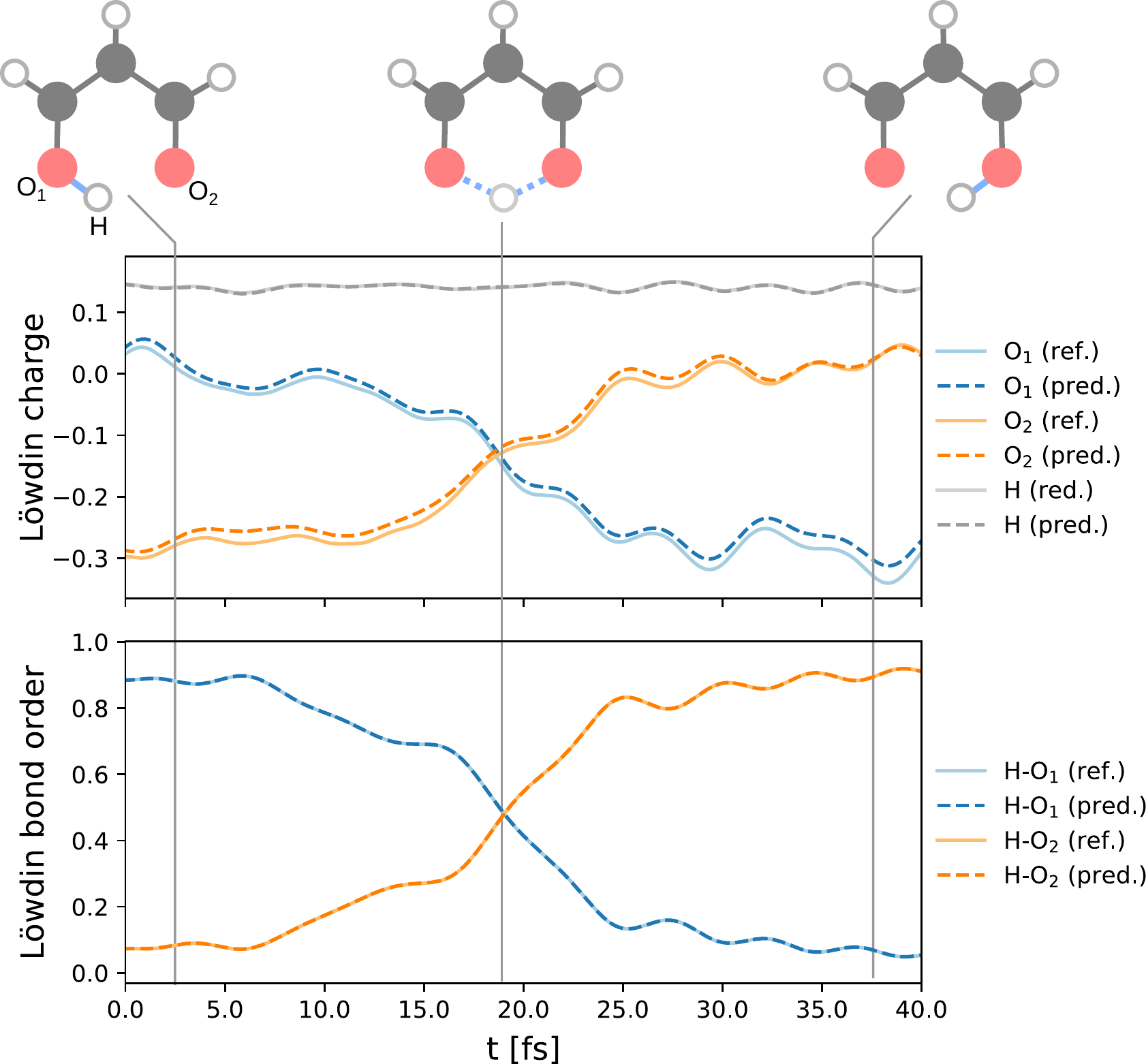}
	\caption{Population analysis for the proton transfer event in the enol form of MDA using the QUAMBO transformed density matrix. The top panel depicts the evolution of L{\"o}wdin partial charges of selected atoms based on the reference data (solid lines) and the SchNOrb prediction (dashed lines). The bottom panel visualizes the change in bond orders.}
	\label{fig:results_quambo_pop}
\end{figure}

SchNOrb directly predicts the Hamiltonian in local orbital representation and thereby offers access to numerous properties that can be formulated as expectation values of the molecular wave functions or that are directly related to the density matrix.~\cite{Schutt2019} This includes important chemical interpretation tools such as atomic partial charges and bond orders that offer insights into the local electronic structure. Such information is not accessible with total-energy-based ML models. This is true for SchNOrb predictions in full basis and in QUAMBO representation.

As an example, Fig~\ref{fig:results_quambo_pop} depicts SchNOrb predicted L{\"o}wdin charges and bond orders in QUAMBO representation for the proton transfer process in MDA.
According to the population analysis, charge is transferred between the two oxygens involved in the reaction, as their partial charges exchange sign along the reaction path.
Interestingly, the transferred hydrogen only shows minimal charge fluctuations compared to the oxygen atoms.
This indicates, that the electronic density moves along with the proton, which experiences the same electronic environment during the process. This is to be expected during an adiabatic hydrogen transfer.
A similar result was found in the original SchNOrb publication, where we presented the associated change in electron density as predicted by full basis SchNOrb.~\cite{Schutt2019}
It is also possible to directly follow the formation and creation of the oxygen-hydrogen bonds using the concept of bond orders (bottom panel of Fig.~\ref{fig:results_quambo_pop}).
The prediction errors associated with partial charges and bond orders for full basis and QUAMBO SchNOrb are reported in Supplementary Tab.~2 of the SI. The associated prediction errors are reduced by several orders of magnitude in the QUAMBO representation when compared to full basis.
These demonstrations highlight the benefits of encoding electronic structure information into machine learning models in general and SchNOrb in particular, which yields accurate predictions for both bond orders and partial charges.

Lastly, combining SchNOrb with QUAMBO representations also improves the computational efficiency of the models. In the case of aspirin, training times are reduced from 36.7~minutes per epoch (full basis) to 20.9~minutes (QUAMBO) on  a GeForce GTX 1080 GPU. Evaluation times profit in a similar manner, where the full prediction for a single sample (prediction of Hamiltonian and overlap matrices and solution of the eigenvalue problem) takes 45.4~ms with a full basis and only 25.4~ms in the QUAMBO representation. Interestingly, prediction times for the matrices only are similar in both approaches (26.4~ms for the full basis and 23.0~ms for QUAMBOs) and the primary speedup is obtained during the solution of the eigenvalue problem (18.9~ms and 2.3~ms for full basis and QUAMBOs, respectively), where the QUAMBO models profit from the greatly reduced size of the matrices.

\section{\label{discussion}Conclusions and Outlook}

The SchNOrb deep learning model provides a framework for the  analytical parametrization of molecular electronic wavefunctions by representing the Hamiltonian in a predefined local atomic orbital basis. In this work and in our previous publication~\cite{Schutt2019}, we show that SchNOrb accurately reproduces molecular Hamiltonians, their eigenvalues and eigenfunctions, as well as derived properties across configuration space by learning the rotational equivariance properties of basis function pairings via their local chemical environment. This is not done by explicit encoding of the rotational equivariance properties of the underlying basis functions, but by means of rotational data augmentation. This approach does lead to a higher demand for training data and extended training times, but also enables us to train on arbitrary basis functions that do not have canonical angular transformation properties, such as optimally tuned minimal basis functions. The QUAMBO representation is an example for such a basis. It is based on a principal component analysis of virtual space to extract the relevant contributions, which are then mixed with occupied eigenstates to accurately represent the occupied eigenvalue subspace in a quasi-atomic local orbital basis.

Minimal basis representation (one basis function per valence electronic state) is an important aspect of efficient and scalable electronic structure descriptions and is a common feature of many existing semi-empirical and tight-binding approximations in quantum chemistry and electronic structure theory. As shown in this work, the SchNOrb deep learning model can accurately represent DFT Kohn-Sham wavefunctions and energy levels of organic molecules in minimal basis. We believe that this approach has the potential to become a framework to construct highly accurate and transferable tight-binding methods from first principles electronic structure theory. Such models are crucial to simulate complex dynamical properties of molecules that depend on the coupling of electronic and nuclear motion. Possible applications range from dynamical spectroscopy and photochemical simulations to transport through molecular junctions. 

Deep learning based models of electronic wavefunctions have the potential to reshape future quantum chemistry workflows. SchNOrb trains on quantum chemical data and its output can directly feed into quantum chemical calculations, for example as accurate wave function guess for SCF calculations or as starting point for correlated wavefunction calculations such as MP2.~\cite{Schutt2019} There are many further directions to explore with this model, but also a number of limitations that need to be overcome. For example, with larger and larger system size, rotational data augmentation eventually becomes unfeasible, even in minimal basis representation. Therefore some form of explicit encoding of the translational and rotational equivariance properties of the wavefunctions within the model will become crucial to ensure consistent model performance. Something to that effect has recently been proposed for 3D point clouds, which likely is adaptable to local spherical harmonic basis functions.~\cite{Thomas2018} Equally, while SchNOrb provides access to many properties which aid chemical interpretation and conceptual explanation, for example bond orders and atomic partial charges, the model parameters inside the deep learning model of the Hamiltonian do not provide much interpretational value. Therefore, there is clearly further scope for the development of more explainable and interpretable ML representations of quantum mechanical wavefunctions. 

In conclusion, the SchNOrb method can accurately parametrize molecular wavefunctions of large organic molecules if provided with a suitable basis representation. Here we have used the quasi-atomic minimal basis orbitals or QUAMBOs, which also showcases the ability of SchNOrb to train on basis functions which are not pure angular momentum functions. 

\section*{Data Availability Statement}
The training data  and code base used in this study are available from the authors upon reasonable request.

\begin{acknowledgments}
	
	This project has received funding from the European Unios Horizon 2020 research and innovation program under the Marie Sk\l{}odowska-Curie grant agreement No. 792572 (M.G.).
	A.MS. acknowledges support by the University of Warwick Research Development Fund and a research project grant by the EPSRC-funded AI3SD Network+ (EP/S000356/1). M.L. acknowledges support by the EPSRC Centre for Doctoral Training of Mathematics for Real-World Systems (EP/L015374/1). R.J.M. acknowledges funding by the UKRI Future Leaders Fellowship programme (MR/S016023/1). The authors acknowledge fruitful discussions with Klaus-Robert M\"uller (TU Berlin). Computing resources have been provided by the Scientific Computing Research Technology Platform of the University of Warwick, the EPSRC-funded HPC Midlands+ computing centre (EP/P020232/1), and the EPSRC-funded high-end computing Materials Chemistry Consortium (EP/R029431/1). Correspondence to M.G. and R.J.M.
\end{acknowledgments}


\nocite{*}
\bibliography{gastegger_quambo_schnorb}

\clearpage
\onecolumngrid
\appendix

\renewcommand{\thefigure}{S\arabic{figure}}
\renewcommand{\thetable}{S\arabic{table}}
\setcounter{figure}{0}

\section*{Supplementary Material}

\captionsetup[figure]{justification=raggedright}
\captionsetup[table]{justification=raggedright}

\begin{figure}[h!]
	\includegraphics[width=1.0\columnwidth]{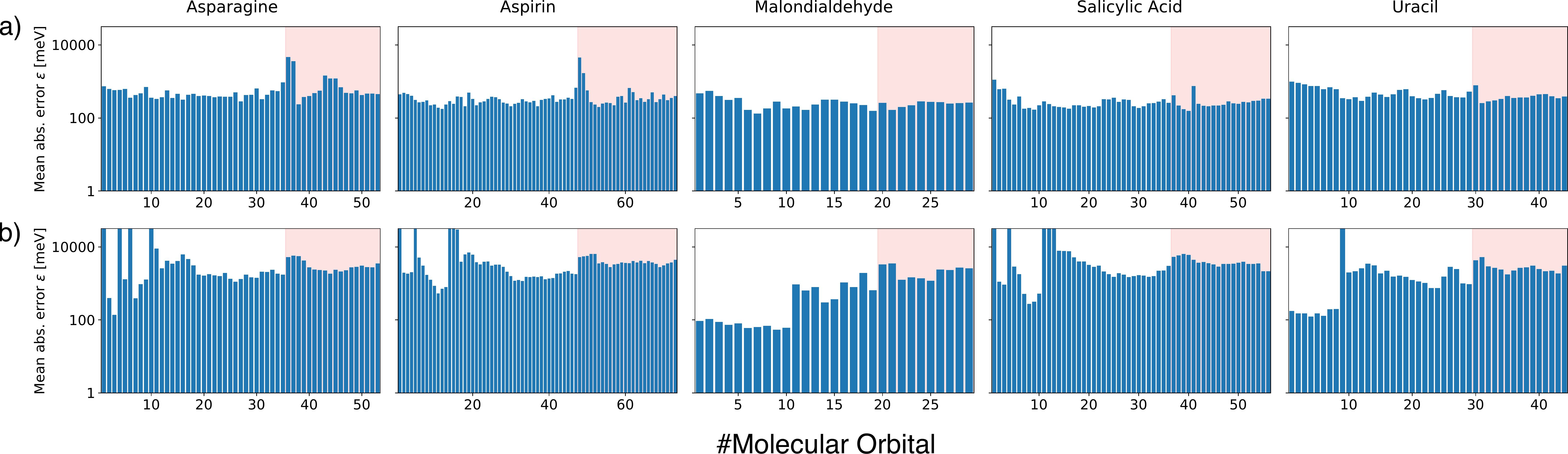}
	\caption{Maximum absolute test errors associated with the molecular orbital energies predicted by SchNOrb models a) using the QUAMBO pre-processing step and b) the full basis set.  In In case of the latter, the same number of energy levels as obtained for the QUAMBO basis were included in the plot for reasons of clarity.  A red background indicates unoccupied orbital space. \label{sfig:max_erros_eval}}	
\end{figure}

\begin{figure}[h!]
	\includegraphics[width=1.0\columnwidth]{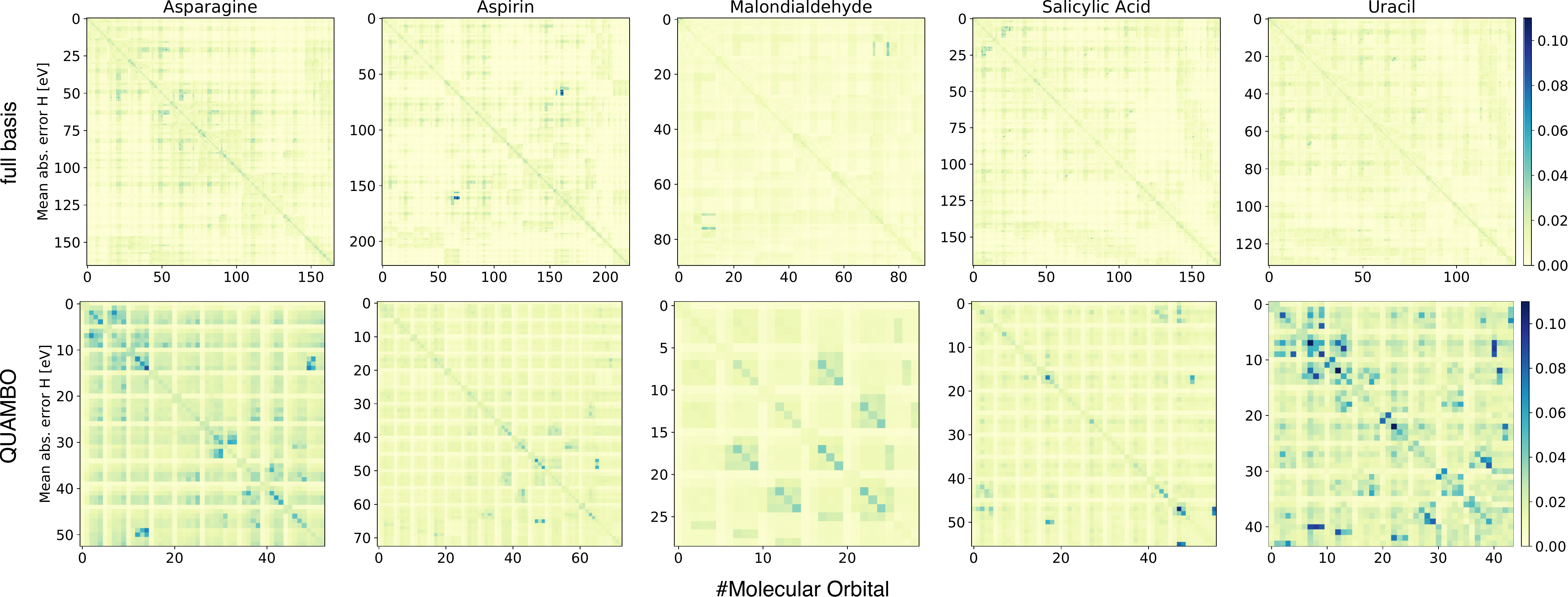}
	\caption{Mean absolute test error for the individual elements of the full (left) and QUAMBO transformed Hamiltonian (right) for all studied molecules. \label{sfig:h_errors}}
\end{figure}

\begin{figure}[h!]
	\includegraphics[width=1.0\columnwidth]{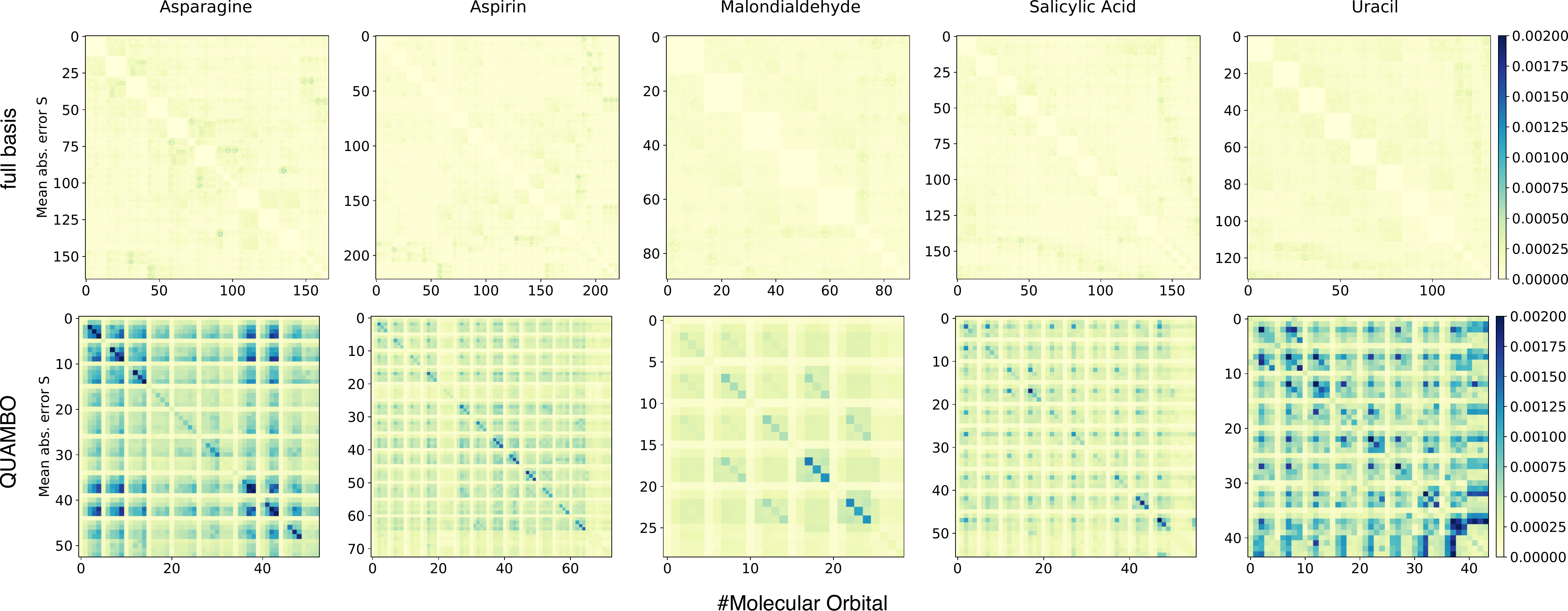}
	\caption{Mean absolute test error for the individual elements of the full (left) and QUAMBO transformed overlap matrices (right) for all studied molecules. \label{sfig:s_errors}}
\end{figure}

\begin{figure}[h!]
	\includegraphics[width=1.0\columnwidth]{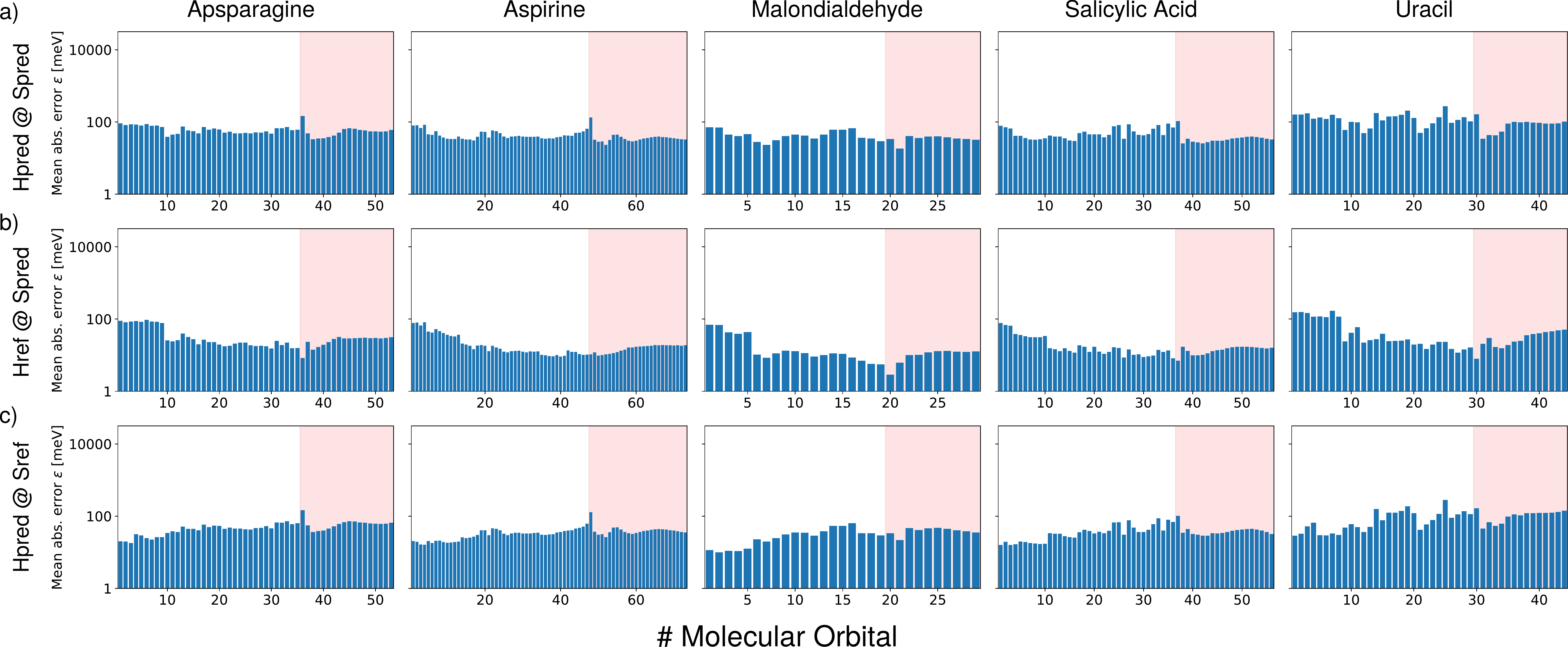}
	\caption{Mean absolute test errors of the molecular orbital energies predicted by QUAMBO-SchNOrb models using a) predicted Hamiltonians and overlap matrices, b) reference Hamiltonianis and predicted overlaps and c) predicted Hamiltonians and reference overlaps for the computation of the eigenvalues. \label{sfig:s_error_components}}
\end{figure}

\begin{table}
	\caption{Train, validation and test set splits, as well as sizes of the mini batches used to train the models. The different split sizes are due to the different convergence behavior of SchNOrb models trained on full basis or QUAMBO matrices. Small differences in the overall data set sizes are due to the removal of data points for which the QUAMBO transformation did not converge.\label{stab:training}}
	\centering
	\begin{tabular}{lccccccccc}
		\toprule
		\multirow{2}{*}{Dataset} & \multicolumn{4}{c}{Full Basis} && \multicolumn{4}{c}{QUAMBO} \\
		\cmidrule{2-5} \cmidrule{7-10}  
		& train  & validation & test & batch && train  & validation & test  & batch\\
		\midrule
		Asparagine      & 25000 & 500 & 4500 & 32 && 20000 & 2000 & 7999 & 20 \\
		Aspirin         & 25000 & 500 & 4500 & 32 && 20000 & 2000 & 7997 & 10 \\ 
		Malondialdehyde & 25000 & 500 & 1478 & 32 && 20000 & 2000 & 4978 & 20 \\
		Salicylic Acid  & 25000 & 500 & 4500 & 36 && 20000 & 2000 & 7999 & 20 \\
		Uracil          & 25000 & 500 & 4500 & 48 && 20000 & 2000 & 8000 & 20 \\
		\bottomrule
	\end{tabular}
\end{table}

\begin{table}
	\caption{Mean absolute test errors on atomic charges and bond orders as predicted by SchNOrb models trained on a full basis (full) and QUAMBO representation (QUAMBO).\label{stab:pop}}
	\centering
	\begin{tabular}{lrrrrrrrrrrr}
		\toprule
		\multirow{2}{*}{Dataset} & \multicolumn{2}{c}{Mulliken charge} && \multicolumn{2}{c}{L{\"o}wdin charge} && \multicolumn{2}{c}{Mulliken bond order} && \multicolumn{2}{c}{L{\"o}wdin bond order} \\
		\cmidrule{2-3} \cmidrule{5-6}  \cmidrule{8-9}  \cmidrule{11-12} 
		& full  & QUAMBO    && full  & QUAMBO   && full  & QUAMBO  && full & QUAMBO\\
		\midrule
		Asparagine      & 0.117 & 0.004 && 0.056 & 0.004 &&   7.646 & 0.002 && 0.005 & 0.001 \\
		Aspirin         & 0.307 & 0.004 && 0.171 & 0.004 && 290.351 & 0.001 && 0.018 & 0.001 \\
		Malondialdehyde & 0.030 & 0.004 && 0.004 & 0.004 &&   0.180 & 0.002 && 0.001 & 0.001 \\
		Salicylic Acid  & 0.267 & 0.005 && 0.136 & 0.005 && 121.136 & 0.001 && 0.025 & 0.001 \\
		Uracil          & 0.122 & 0.010 && 0.047 & 0.010 &&   8.175 & 0.003 && 0.006 & 0.002 \\
		\bottomrule
	\end{tabular}
\end{table}

\begin{table}
	\caption{Mean absolute test errors on total energies as predicted by SchNOrb models trained on a full basis (full) and QUAMBO representation (QUAMBO).\label{stab:ener}}
	\centering
	\begin{tabular}{lrr}
		\toprule
		\multirow{2}{*}{Dataset} & \multicolumn{2}{c}{Energy [eV]} \\
		\cmidrule{2-3}
		& full  & QUAMBO \\
		\midrule
		Asparagine      & 0.001182 & 0.021366 \\
		Aspirin         & 0.001428 & 0.016935 \\
		Malondialdehyde & 0.000353 & 0.004668 \\
		Salicylic Acid  & 0.000821 & 0.010575 \\
		Uracil          & 0.000848 & 0.017600 \\
		\bottomrule
	\end{tabular}
\end{table}

\begin{table*}
	\centering
	\caption{Mean absolute test errors for two QUAMBO-SchNOrb models of malondialdehyde using a different tradeoff $\eta$ for the total energy term in the loss function during training. Reported are values for the Hamiltonians $H$, overlap matrices $S$, orbital energies $\epsilon$, HOMO-LUMO gap (Gap) and the total energy E$_\mathrm{tot}$.}
	\label{stab:comparison}
	
	\begin{tabular}{lrrrrrrrrrr}
		\toprule
		& H [eV]  && S  && $\epsilon$ [eV]   && Gap [eV]  && E$_\mathrm{tot}$ [eV] \\
		\midrule
		$\eta=0.0001$ &  0.009474 && 0.000213 && 0.0449 && 0.0422 && 0.004668 \\
		$\eta=1.0$    &  0.008676 && 0.000202 && 0.0220 && 0.0282 && 0.000296 \\
		\bottomrule
	\end{tabular}
	
\end{table*}

\end{document}